\documentclass[10pt,twocolumn,letterpaper]{article}

\usepackage{cvm}
\usepackage{times}
\usepackage{epsfig}
\usepackage{graphicx}
\usepackage{amsmath}
\usepackage{amssymb}

\usepackage[pagebackref=true,breaklinks=true,letterpaper=true,colorlinks,bookmarks=false]{hyperref}

\cvmfinalcopy 

\def\cvmPaperID{} 

\ifcvmfinal\fi

\begin{document}

\title{Automated pebble mosaic stylization of images}

\author{Lars Doyle\\
{\tt\scriptsize lars.doyle@carleton.ca,}
\and
Forest Anderson\\
{\tt\scriptsize forest.anderson@carleton.ca}
\and
Ehren Choy\\
{\tt\scriptsize ehren.choy@gmail.com}
\and
David Mould\\
{\tt\scriptsize mould@scs.carleton.ca}
\and 
Carleton University, Ottawa, Canada \\
}

\maketitle

\begin{abstract}
Digital mosaics have usually used regular tiles, simulating the historical ``tessellated'' mosaics. In this paper, we present a method for synthesizing \emph{pebble mosaics}, a historical mosaic style in which the tiles are rounded pebbles. We address both the tiling problem, where pebbles are distributed over the image plane so as to approximate the input image content, and the problem of geometry, creating a smooth rounded shape for each pebble. We adapt SLIC, simple linear iterative clustering, to obtain elongated tiles conforming to image content, and smooth the resulting irregular shapes into shapes resembling pebble cross-sections. Then, we create an interior and exterior contour for each pebble and solve a Laplace equation over the region between them to obtain height-field geometry. The resulting pebble set approximates the input image while presenting full geometry that can be rendered and textured for a highly detailed representation of a pebble mosaic.
\end{abstract}

\section{Introduction}\label{sec:introduction}

Mosaics are an art form that dates back thousands of years.
The earliest historical mosaics were {\em pebble mosaics}~\cite{Dunbabin:1999:MGR,Ling:1998:AM},
whose component pebbles were heterogeneous in size and shape.
Pebble mosaics were floors paved with pebbles, where the
pebbles were arranged so as to form an image or design.
The craft of pebble mosaics continues into the
21st century~\cite{Howarth:2003:CPM} with new pebble mosaics 
being built by hobbyists and city planners.

\begin{figure}[htbp]
\begin{center}
   \includegraphics[width=1\linewidth]{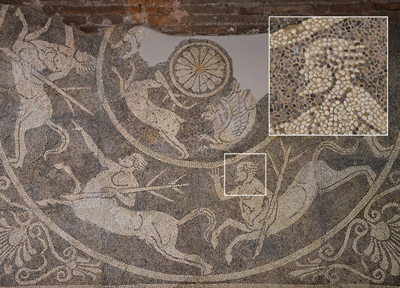}
\end{center}
   \caption{Fragment of a pebble mosaic floor dating from the 4th century BCE}
\label{fig:ancient}
\end{figure}

Pebble mosaics, as well as the contemporaneous {\em chip mosaics}
made of fragments of quarried stone~\cite{Dunbabin:1999:MGR},
use entirely irregular tiles. 
The archetypal mosaic
is the {\em tessellated} mosaic, made of regular cubes of stone
(tesserae). 
 The tessellated mosaics are most familiar to us
and have been the most thoroughly studied in computer graphics.
Tessellated mosaics have been dated to the third century BCE. However,
pebble mosaics appeared in Greece
hundreds of years earlier~\cite{Dunbabin:1999:MGR} and
have not received much attention in computer graphics.
In this paper, we propose a novel algorithm for irregular pebble mosaics,
using a variant of SLIC~\cite{Achanta:2012:SLIC} to obtain an initial segmentation, smoothing the resulting boundaries, and using a Poisson solver to interpolate a smooth heightfield for each pebble which we can then render using conventional lighting and texturing.

\begin{figure*}[htbp]
\begin{center}
   \includegraphics[width=0.19\linewidth]{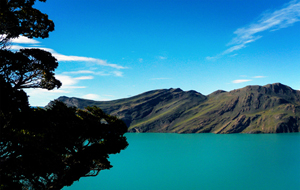}
   \includegraphics[width=0.19\linewidth]{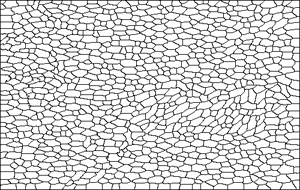}
   \includegraphics[width=0.19\linewidth]{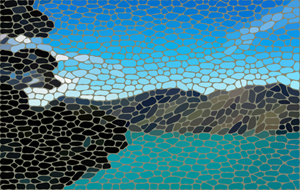}
   \includegraphics[width=0.19\linewidth]{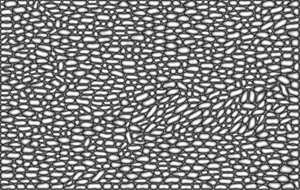}
   \includegraphics[width=0.19\linewidth]{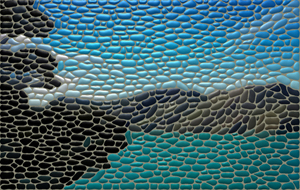}
\end{center}
   \caption{An image progressing through our system from left to right: input, segmentation, boundary smoothing, pebble geometry, lighting.}
\label{fig:process}
\end{figure*}

For a mosaic to successfully convey an image, it is important to align
tile edges with image edges. The use of square tiles
imposes severe restrictions on the detail level that can be captured;
our irregular tiles can convey considerable details, including interior
edges of figures, something often neglected in previous techniques.
Our algorithm is entirely automatic;
users can optionally guide the output
by annotating the input image with an importance map or manually adding decorative features in a preprocessing phase.

This paper makes two main contributions. First, we adapt SLIC so that it is suitable for creating irregular, elongated pebble shapes. We estimate the local direction of the image and then bias the SLIC clustering distance according to a local coordinate system, producing natural-looking size and aspect ratio variations. Second, we compute smooth pebble geometry for the resulting tiles. We use the Laplace equation, 
setting up constraints and then solving to meet them, thus producing smooth shapes resembling river pebbles. By creating and rendering this geometry, we bridge photorealism and non-photorealism.

The paper is organized as follows. Section~\ref{pw} reviews previous
works on computer-generated mosaics. Section~\ref{algo} describes
our algorithm in detail. Section~\ref{results} shows images created
using our method and discusses its benefits and drawbacks. Finally,
Section~\ref{con} summarizes the work and suggests future directions.

\section{Related work} \label{pw}

Battiato et al.~\cite{Battiato:2007} propose a taxonomy of digital mosaic research where the two initial branches divide {\em tile} mosaics from {\em multi-picture} mosaics. This distinction stems from the nature of the basic picture elements. In tile mosaics, the image plane is divided into small regions, each individually colored to represent the underlying input image. In contrast, multi-picture mosaics employ a dataset of images that are used to assemble an approximation to the input image based on local color and structure similarity; the typical result is a photomosaic~\cite{Silvers:1997}. We situate our current work within the tile mosaic branch.

In the seminal ``Paint by Numbers''~\cite{Haeberli:1990:PBN},
Haeberli introduced many of the concepts that have since been used for
mosaic emulation. His idea of using Voronoi diagrams for mosaics
has been used in commercial products and in subsequent research;
Centroidal Voronoi Diagrams (CVD's) are particularly common.
CVD's are often achieved by Lloyd's algorithm,
a relaxation process that repeatedly moves the Voronoi centres to
the centroids of their regions. The CVD process has formed the basis
for considerable work in mosaic and stipple
creation~\cite{Hausner:2001:SDM,Secord:2002:WVS,Hiller:2003:BS},
since it is a good way to distribute points on the plane.

Hausner~\cite{Hausner:2001:SDM} presented
 an iterative algorithm for placing mosaic
centres, using hardware-accelerated
CVD's to distribute tiles.
Hausner also identified a crucial issue in mosaics:
that tile edges should be aligned with image edges.
Hausner resolved this in his work
by having tiles move away from user-specified edges.
An alternative method for achieving edge alignment was given
by Elber and Wolberg~\cite{Elber:2003:RTM},
who arrange rows of tiles along streamlines
parallel to initial user-specified curves.
Yet another way of addressing edge alignment was given by
 Di Blasi and Gallo~\cite{DiBlasi:2005:AM},
who propose to cut the rectangular tiles where they
 cross image edges.
Liu et al.~\cite{Liu:2010:CGM} use graph cuts rather than explicit edge detection
to prevent tiles from crossing image edges.

Within the multi-picture mosaic branch a thread of research involves populating a set of container shapes with tiles, generally without any intention of providing interior image detail. Kim and Pellacini's
Jigaw Image Mosaic~\cite{Kim:2002:JIM} is an example, where the method
produces an irregular tiling of the
image plane with predefined tiles,
minimizing a set of error criteria including
tile overlap and color mismatch. 
More recent work by Saputra et al.~\cite{flowpak, repulsionpak}
arranges figures within the container shape while seeking an
aesthetic distribution rather than a full packing. Kwan et al.~\cite{Kwan:2016} accelerate partial-shape matching, through their {\em pyramid of arclength descriptor}, for packing irregular shapes.

Other methods for distributing primitives and tiling the plane
have been devised, and we briefly mention a few others.
Smith et al.~\cite{Smith:2005:A} focussed on coherent movement of
tiles to create animated mosaics; later,
Dalal et al.~\cite{Dalal:2006:SAN} used Fourier transforms to find
good packings of input primitives.
 Kaplan and Salesin~\cite{Kaplan:2000:E,Kaplan:2004:DE}, worked on automatically controlling tile shapes to produce Escher-like tilings where the tiles were close to an input goal shape. Similarly, Goferman et al.~\cite{Goferman:2010} extract irregular regions of interest from a series of photographs and pack them in a puzzle-like manner within a chosen aspect ratio. Photo collage is a related area, but removes the constraint that an underlying image or containing shape must be represented. Using convolutional neural networks, Liu et al.~\cite{Liu:2018} produce photo collages by grouping together images with similar content over the image plane.

\section{Constructing Pebble Mosaics} \label{algo}

\begin{figure}[htbp]
\begin{center}
   \includegraphics[width=0.85\linewidth]{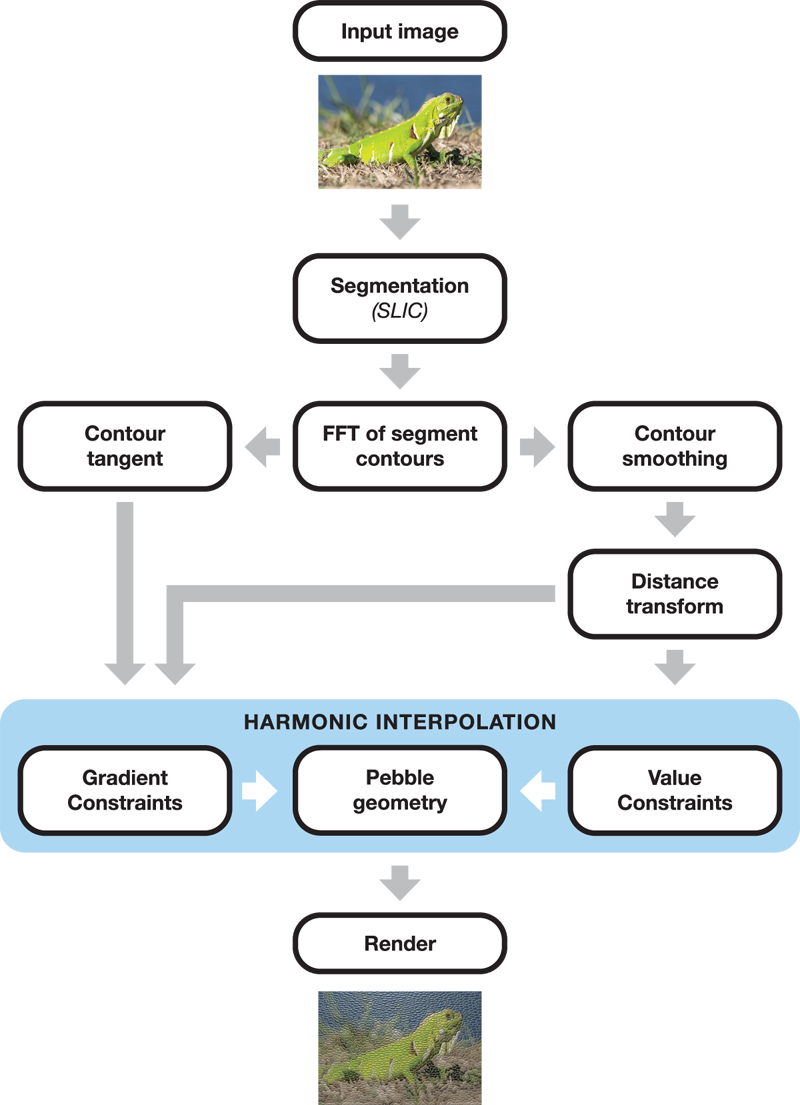}
\end{center}
   \caption{Pipeline of our proposed method.}
\label{fig:pipeline}
\end{figure}

In our approach, we tile the image plane using heterogeneous, 3D pebble-shaped objects. As in previous methods~\cite{Hausner:2001:SDM,Elber:2003:RTM,DiBlasi:2005:AM}, our tiles avoid crossing image boundaries and are oriented to align with a direction map. However, we take a different approach towards this goal. Section~\ref{seg} describes how we modify the {\em Simple linear iterative clustering} algorithm (SLIC)~\cite{Achanta:2012:SLIC} to produce oriented pebble shapes. We take advantage of the inherent boundary-avoiding quality of SLIC and thus have no need for explicit edge detection nor associated parameters or thresholds. We describe how we simplify the boundaries of the initial segmentation in Section~\ref{smooth} to produce smooth, `river-worn' pebbles. Finally, in Section~\ref{3D} we construct a heightfield from the 2D boundaries to extend pebbles into 3D, applying lighting to the resulting geometry. A schematic representation of our algorithm pipeline shows how an input image $I$ is transformed into a pebble mosaic in Figure~\ref{fig:pipeline}. 

\begin{figure}[htbp]
\begin{center}
   \includegraphics[width=0.45\linewidth]{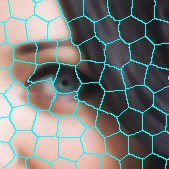}
   \includegraphics[width=0.45\linewidth]{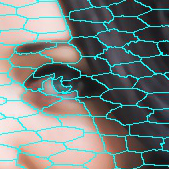} \\[0.05cm]
   \includegraphics[width=0.45\linewidth]{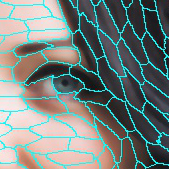}
   \includegraphics[width=0.45\linewidth]{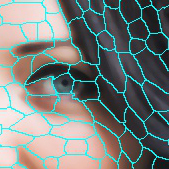}
\end{center}
   \caption{Top left: original SLIC; top right: scaling $v_y$ in Equation~\ref{SLIC_xy} by $\alpha = 3$; bottom left: scaling $\vec{v} \cdot \vec{b_1}$ in Equation~\ref{SLIC_spatial} by $\alpha = 3$; bottom right: using random scaling in Equation~\ref{SLIC_spatial}.}
\label{fig:SLIC}
\end{figure}

\subsection{Segmentation} \label{seg}

SLIC produces compact super-pixels by clustering pixels into groups, based on colour and spatial distance. Their tendency to adhere to image boundaries is beneficial for describing image content and forms the basis of our pebble shapes. In its original formulation, the spatial distance of a pixel $p$ from a cluster center $c$ can be described by an offset vector $\vec{v} = p-c$, allowing us to compute the $l_2$ distance as:
\begin{equation}\label{SLIC_xy}
D_s = \sqrt{ v_x^2 +  v_y^2},
\end{equation}
where $v_x$ and $v_y$ are the components of $\vec{v}$ parallel to the $x$ and $y$-axes. It is often the case in nature that pebbles are longer in one dimension than the other, forming approximate, oval-like boundaries as opposed to circular. As a first modification to Equation~\ref{SLIC_xy} we can apply different scaling factors to the $x$ and $y$ components of $\vec{v}$. This results in the elongated super-pixels that are shown in the top right image of Figure~\ref{fig:SLIC}.

Artists often take advantage of pebble shapes and will emphasize image edges by aligning the long side of a pebble parallel to an edge. We can approximate this effect with one further modification to our distance metric. First, we construct a structure tensor~\cite{Brox:2006} at each super-pixel center by integrating the matrix field, $\nabla I \nabla I^T$, weighted by a Gaussian function. The tensor's unit eigenvectors $e_1$ and $e_2$, associated with eigenvalues $\lambda_1 \geq \lambda_2$, point parallel and perpendicular to the smoothed image gradient. We can now use these vectors as a new basis in our distance calculation. Furthermore, applying a larger weight to the component of $\vec{v}$ parallel to $e_1$ than $e_2$ will allow super-pixels to spread tangent to image edges. This effect can be seen in Figure~\ref{fig:SLIC} on the bottom left. In flat or corner regions, where there is inadequate orientation information, we simply assign a default direction. This assessment is made by thresholding an orientation coherence estimate:
\begin{equation}\label{coherence}
C = \sqrt{\frac{\lambda_1 - \lambda_2}{\lambda_1 + \lambda_2 + K}},
\end{equation}
where $K$ is a constant chosen to avoid division by zero and to de-emphasize weak tensors.

The final distance metric is:
\begin{equation}\label{SLIC_spatial}
D_s = \sqrt{\alpha_1 \vec{v} \cdot \vec{b_1} + \alpha_2 \vec{v} \cdot \vec{b_2}},
\end{equation}
where $\alpha_1$ and $\alpha_2$ are scaling factors, controlling both the aspect ratio and the overall size of each cluster. The vectors $\vec{b_1}$ and $\vec{b_2}$ correspond either to the local image orientation, if there a strong local orientation exists, or a default direction. The decision is made by comparing $C$ to a threshold $T_{coh}$ as follows:
\begin{equation}\label{basis}
\vec{b_1} = \begin{cases} \vec{e_1} & C > T_{coh} \\ \vec{d_1} & C \leq T_{coh} \end{cases} \text{ and }
\vec{b_2} = \begin{cases} \vec{e_2} & C > T_{coh} \\ \vec{d_2} & C \leq T_{coh} \end{cases}.
\end{equation}
The vectors $d_1$ and $d_2$ comprise a default orthonormal basis. In our examples we set $T_{coh}$ to 0.5 and $d_1$ to the y-axis.

\begin{figure}[t]
\begin{center}
   \includegraphics[width=0.24\linewidth]{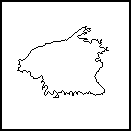}
   \includegraphics[width=0.24\linewidth]{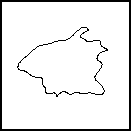}
   \includegraphics[width=0.24\linewidth]{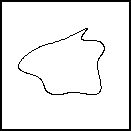}
   \includegraphics[width=0.24\linewidth]{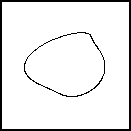}
\end{center}
   \caption{Left: original contour; remaining images: reconstructing the contour from $L =$ 37, 17, and 7 Fourier coefficients.}
\label{fig:contour_smoothing}
\end{figure}

The scaling factors, $\alpha_1$ and $\alpha_2$ are selected individually for each super-pixel guided by a random process, such that
\begin{equation}\label{alph_scale}
 \alpha_1= \phi_{a1}\phi_{s} \text{ and } \alpha_2= \phi_{a2}\phi_{s}. 
 \end{equation}
Through experimentation, we chose to compress the aspect ratio perpendicular to edges by $\phi_{a1}=3$. The other terms are determined by two uniform random numbers $r_1, r_2 \in [0,1]$. We then set $\phi_{a2}=(\phi_{a1}-1)r_1^2 + 1$ and set the scale term $\phi_{s} = r_2^2 +1$. 

The local distance metric $D_s$ is used in the SLIC process to oversegment the image. We refer to the resulting oversegmented image as $P$, and each segment, $P_i \in P$, is a {\em pebble}.

\begin{figure}
\begin{center}
   \includegraphics[width=0.48\linewidth]{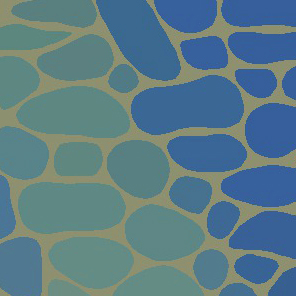}
   \includegraphics[width=0.48\linewidth]{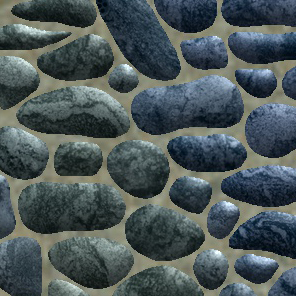}
\end{center}
   \caption{High-resolution pebbles rendered at 5 times the input resolution; left: pebble shapes; right: 3D rendered pebbles.}
\label{fig:upsampling}
\end{figure}
\begin{figure}
\begin{center}
   \includegraphics[width=0.40\linewidth]{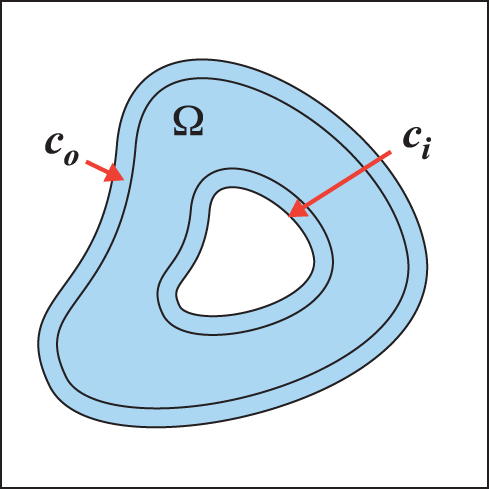}
   \includegraphics[width=0.40\linewidth]{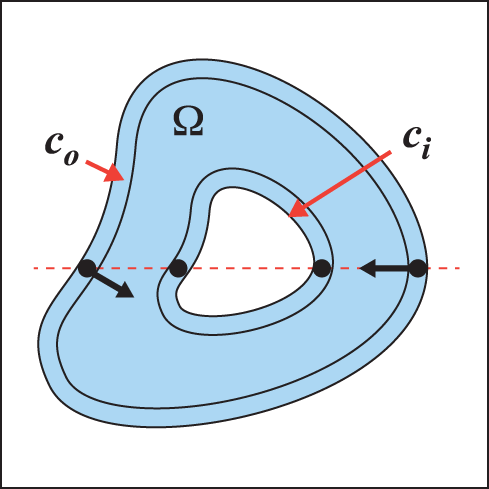}
\end{center}
\caption{Left: the domain, $\Omega$, and boundaries ($c_o$ and $c_i$) of $P_i$; right: the gradient orientation on $c_o$ (arrows) and zero-gradient on $c_i$ (dots).}
\label{fig:domain}
\end{figure}

\subsection{Boundary Smoothing}  \label{smooth}

The pebbles that we constructed in Section~\ref{seg} will contain many irregularities that depart from the smooth pebble shapes that we wish to create. Hence, we apply a low-pass filter in the frequency domain~\cite{Gonzalez:2008:DIP}, to each pebble's outer contour $c_o(k)$, for $k = 0,1, \ldots, K-1$. This process effectively reconstructs a contour from $L$ Fourier coefficients, where $L$ is less than $K$. In Figure~\ref{fig:contour_smoothing}, we illustrate a contour reconstructed with various values of $L$. Note that at this stage we can obtain a resolution-independent, tiled image $P$ by applying a scale factor to the Fourier coefficients, obtaining a larger (or smaller) $c_o$ as needed. As seen in Figure~\ref{fig:upsampling}, one advantage of rendering at a higher resolution is the increased surface area that can be used for adding texture and lighting. Finally, each pebble $P_i \in P$ is updated by flood filling its corresponding $c_o$.

\subsection{Pebble geometry}  \label{3D}

We construct a heightfield for each pebble by means of harmonic interpolation over the domain, $\Omega$, that resides between two contours (Figure~\ref{fig:domain}, left). The outer contour, $c_o$, is described above. We obtain the inner contour, $c_i$, by thresholding the normalized distance transform of $P_i$ by $T_{dist} \in (0,1)$. We set a zero gradient at the inner contour, thus creating a small flat face to each pebble which then curves downwards to the image plane.
 In all examples, we set $T_{dist}=0.85$.

Our heightfield is the solution to the Laplace equation~\cite{Perez:2003}:
\begin{equation}\label{poisson}
\Delta P_i = 0 \text{ over } \Omega,
\end{equation}
with boundary value constraints $P_i|_{c_o} = 0$ and $P_i|_{c_i} = 1$. Additionally, we set gradient constraints at the boundaries such that $|\nabla P_i| = 0$ on $c_i$. The gradient on $c_o$ is constructed as follows. Returning to the Fourier transform of Section~\ref{smooth}, we note that the derivative $c_o{'}(k)$ of the sampled function $c_o(k)$ can be computed in the Fourier domain. This process provides us with a sequence of vectors that are tangent to the curve; one for each sample point. Rotating each vector $90^{\circ}$ inwards gives us a gradient orientation that is orthogonal to the boundary. The gradient magnitude is chosen as follows:
\begin{equation}\label{grad_mag}
|\nabla P_i| = \frac{\beta}{T_{dist}D_{max}}, 
\end{equation}
where $D_{max}$ is the maximum value of the distance transform. The parameter $\beta$ determines the shape of the resulting pebble and various settings are illustrated in Table~\ref{fig:poisson}. We choose $\beta = 2$ to construct the pebble profile curving downward into the surrounding area in our examples. Notice that setting $\beta$ too high will result in the gradient overshooting its target at the inner contour resulting in a depression at the center as seen in the bottom row of Table~\ref{fig:poisson}.

\begin{table}
\begin{center}
\begin{tabular}{|c|c|c|c|}
\hline
{\small $\beta$} & {\small Heightfield} & {\small Cross-section} & {\small 3D plot} \\
\hline
\hline
{\small $\beta = 1$} &
\includegraphics[width=0.1\linewidth]{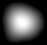} & 
\includegraphics[width=0.25\linewidth]{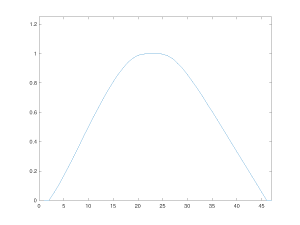} &
\includegraphics[width=0.25\linewidth]{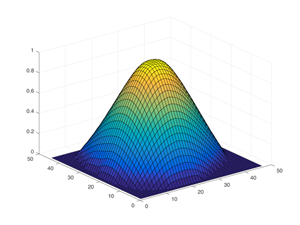} \\
\hline
{\small$\beta = 2$} &
\includegraphics[width=0.1\linewidth]{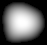} & 
\includegraphics[width=0.25\linewidth]{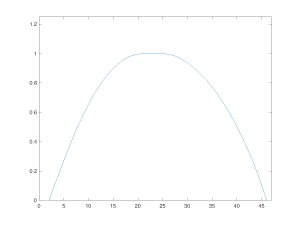} &
\includegraphics[width=0.25\linewidth]{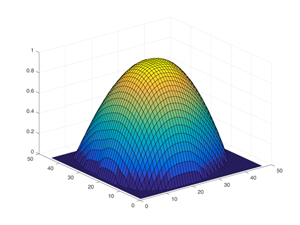} \\
\hline
{\small $\beta = 3$} &
\includegraphics[width=0.1\linewidth]{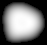} & 
\includegraphics[width=0.25\linewidth]{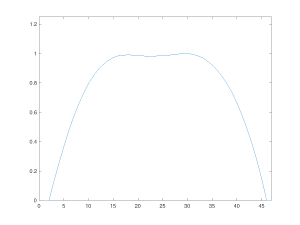} &
\includegraphics[width=0.25\linewidth]{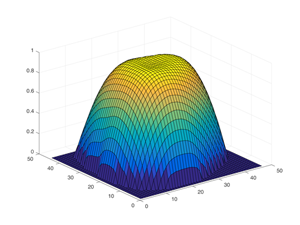} \\
\hline
\end{tabular}
\end{center}
\caption{Constructing a heightfield at varying scales of the gradient magnitude on $c_o$.}
\label{fig:poisson}
\end{table}

\subsection{Rendering} \label{rendering}

\begin{figure}
\begin{center}
   \includegraphics[width=0.85\linewidth]{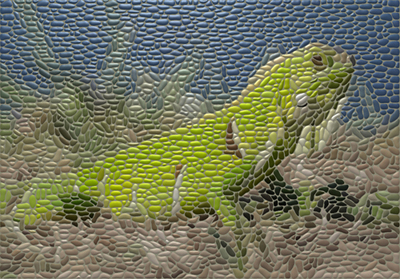} \\[0.05cm]
   \includegraphics[width=0.85\linewidth]{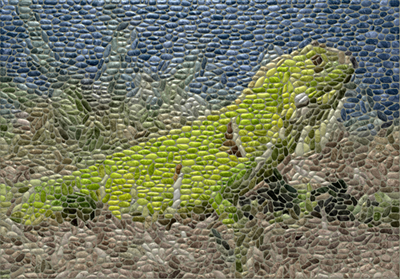}
\end{center}
   \caption{Top: result without texture; bottom: result using texture.}
\label{fig:result}
\end{figure}

We apply Phong shading to the resulting heightfield. We use the average colour in $I$ under $P_i$ as the pebble's surface color. Optionally, we can apply a rock texture to the pebble as well. The texture image, is randomly sampled for each pebble and combined with the luminosity channel using a multiply blend. Example mosaics produced using this scheme are shown in Figure~\ref{fig:result} with and without texture in the top and bottom images, respectively.

\begin{figure*}[t]
\begin{center}
   \includegraphics[width=0.41\linewidth]{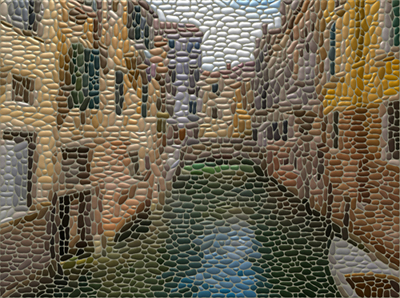}
   \includegraphics[width=0.41\linewidth]{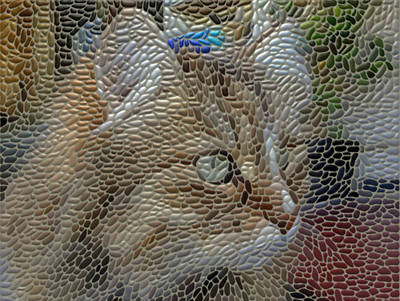} \\[0.05cm]
   \includegraphics[width=0.41\linewidth]{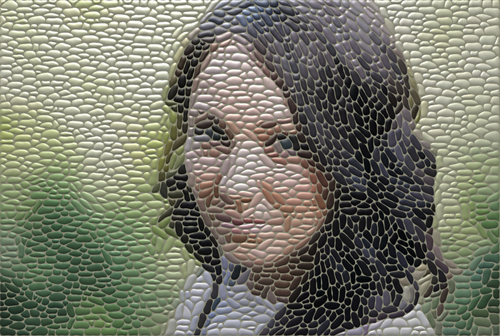} 
   \includegraphics[width=0.41\linewidth]{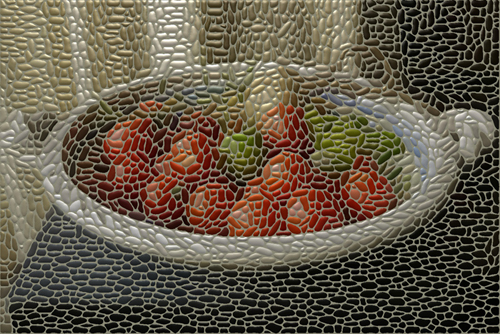} \\[0.05cm]
   \includegraphics[width=0.41\linewidth]{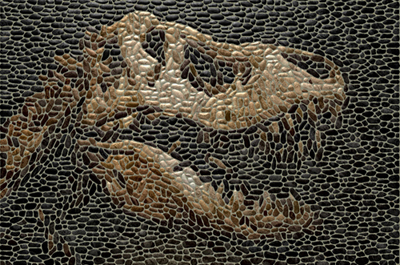} 
   \includegraphics[width=0.41\linewidth]{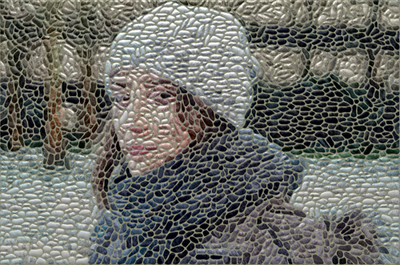} \\[0.05cm]
\end{center}
   \caption{Some results. Top two rows: without texture; bottom row: using marble texture.}
\label{fig:examples}
\end{figure*}

\section{Results and Discussion} \label{results}

We demonstrate our method on photographs containing various subject matter in Figure~\ref{fig:examples}, using 2000 pebbles in each example. The original source images are shown in Figure~\ref{fig:input}. Rendering time for a 1.5 megapixel image is 28 seconds using our unoptimized CPU implementation. The majority of this time is spent solving 2000 $N_i^2$ sparse linear systems in order to construct the geometry of the pebbles. This portion takes 25 seconds of the total 28. Increasing the pebble count leads to solving smaller matrices and thus faster execution times; for example, using 3000 pebbles reduces the solving time to 16 seconds. 

Notice that even at this coarse scale, most of the important image features are still recognizable. The elongated pebble shapes add an impression of motion to the results. This is most noticeable in the cat image at the top left where the pebbles follow the fur orientation. In the portrait image (second row, left) we see how random pebble scaling can add visual interest to otherwise flat image regions. This brings to mind the activity of a mosaicist using tiny pebbles to fill the empty spaces left between larger stones. In the bottom row, adding texture supports the transition from the synthetic 3D shapes in the top rows to a more natural-looking material.

\begin{figure}
\begin{center}
   \includegraphics[width=0.48\linewidth]{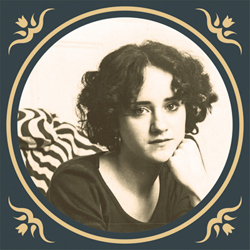}
   \includegraphics[width=0.48\linewidth]{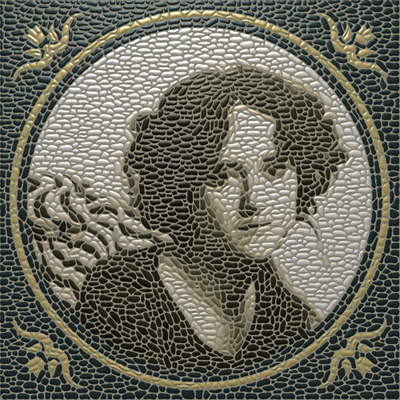} \\[0.05cm]
   \includegraphics[width=0.48\linewidth]{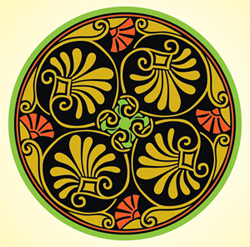}
   \includegraphics[width=0.48\linewidth]{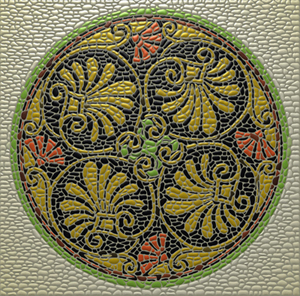}
\end{center}
   \caption{Our method used on ornamental motifs. Left: input images; right: results.}
\label{fig:ornament}
\end{figure}

Inspired by historical mosaics, such as the one depicted in Figure~\ref{fig:ancient}, we demonstrate our method on the ornamental designs shown in Figure~\ref{fig:ornament}. Due to the high contrast in these images, the pebbles adhere well to the image content, creating a striking re-representation of the input.

\subsection{Degrees of Freedom} \label{dof}

Our system has five notable degrees of freedom that can influence the outcome of the final rendered mosaic: color, shape, texture, orientation, and size. We briefly discuss each aspect here.

{\em Color.} Following the tradition in tile mosaics~\cite{Haeberli:1990:PBN, Hausner:2001:SDM, Elber:2003:RTM, DiBlasi:2005:AM} we render each pebble with the average color under the corresponding image region. Alternatively, we could allow color to vary over the pebble region, guided by the input image.

 {\em Shape.} Pebble shape can be influenced by the low pass filter used in the smoothing process discussed in Section~\ref{smooth} and illustrated in Figure~\ref{fig:contour_smoothing}. We chose to retain seven Fourier coefficients, resulting in smooth oval-like pebble shapes. However, less smoothing would provide more shape variety.

 {\em Texture.} We currently limit pebble texture to a single sample but there is potential for more development along this dimension. For example, a database of texture swatches could be employed to match pebble texture with the underlying image. This addition would provide further connection with the input image and increase recognizability.

 {\em Orientation.} Pebbles are oriented parallel to image edges, as is common in both traditional and digital mosaics~\cite{Hausner:2001:SDM, Elber:2003:RTM, DiBlasi:2005:AM}. As described in Section~\ref{seg}, we determine orientation through a structure tensor field, defaulting to a fixed orientation where inadequate information is present. We could also ask the user to provide a vector field in place of a single default direction.  

 {\em Size.} We discuss pebble size in the following subsections, first talking about local variation in pebble dimensions and then discussing size more generally, including the option of varying pebble size based on an importance map.

\begin{figure}
\begin{center}
   \includegraphics[width=0.48\linewidth]{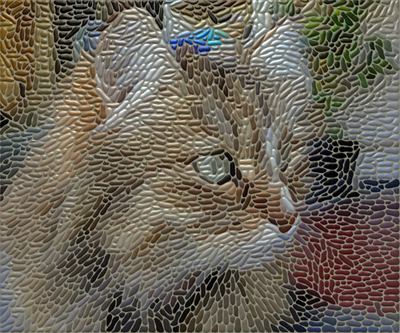}
   \includegraphics[width=0.48\linewidth]{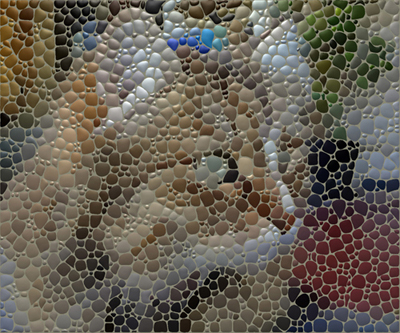} \\[0.05cm]
   \includegraphics[width=0.48\linewidth]{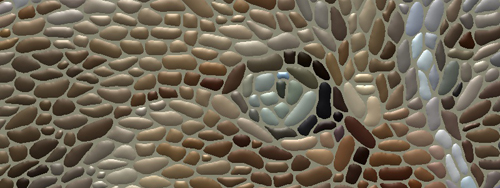}
   \includegraphics[width=0.48\linewidth]{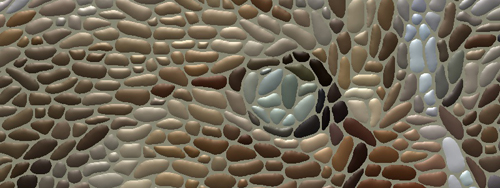}
\end{center}
\caption{Top left: randomly varying the pebble aspect ratio and using a fixed scale; top right: randomly varying the pebble scale and using a fixed aspect ratio; bottom: rendering is nondeterministic due to random scale parameter.}
\label{fig:random}
\end{figure}

\subsubsection{Pebble dimensions} \label{dim}

In Section~\ref{seg} and Equation~\ref{alph_scale} we describe a random process that determines the aspect ratio and relative size of individual pebbles. We now show how varying these parameters can influence the resulting mosaic; the images in the top row of Figure~\ref{fig:random} provide a visual example. On the top left we fix $\phi_{s} = 1$ to maintain a constant size scale and allow the aspect ratio to vary through a randomly generated number. Here we increase $\phi_{a1}$ to 5 and calculate  $\phi_{a2}$ as before. The long thin pebbles work well in this situation where we connect them with the cat's fur. Compare this result to the cat in Figure~\ref{fig:examples}. Here, setting $\phi_{a1}$ to 3 shows less movement in the cat's fur, but a randomly changing $\phi_s$ brings out more variation and liveliness. On the top right, we fix the aspect ratio to $\phi_{a1} = \phi_{a2} = 1$ and allow the scale parameter to vary. We set $\phi_{s} = 5r^2 + 1$, where $r$ is a uniform random number in $[0,1]$. Without orientation information it is more difficult to identify the image. Also, such extreme variability in pebble size is distracting since the sizes are chosen randomly rather than based on image content. On the bottom of Figure~\ref{fig:random} we demonstrate the impact of the random factors in the scaling parameters: note the different outcomes between two runs, using identical parameters, on the left and right. 

\subsubsection{Pebble size} \label{size}

\begin{figure}
\begin{center}
   \includegraphics[width=0.32\linewidth]{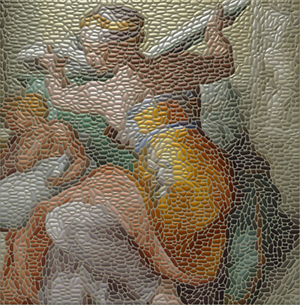}
   \includegraphics[width=0.32\linewidth]{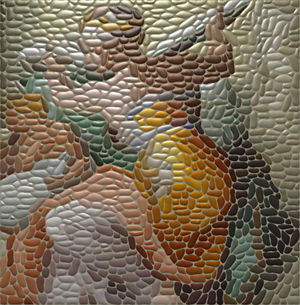}
   \includegraphics[width=0.32\linewidth]{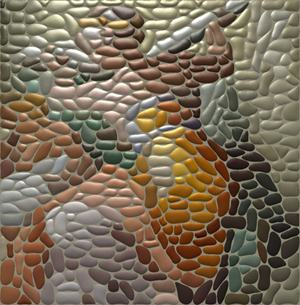}
\end{center}
   \caption{Varying pebble size. Left: 3000 pebbles; center: 1000 pebbles; right: 500 pebbles.}
\label{fig:size}
\end{figure}

In Figure~\ref{fig:size} we vary the number of pebbles that make up a mosaic image. On the left we see a detailed result using 3000 pebbles. Many traditional mosaics, such as the one depicted in Figure~\ref{fig:ancient}, were constructed with this high level of detail. Next, we see a result using 1000 pebbles. Even at this larger size, much of the image remains clear owing to SLIC's tendency to adhere to image boundaries. Finally, the pebble size on the right has probably been pushed too far, making it difficult to recognize the main figure in the result. See Figure~\ref{fig:compare_hausner} for a rendering of this image using 2000 pebbles.

\begin{figure}[t]
\begin{center}
   \includegraphics[width=1\linewidth]{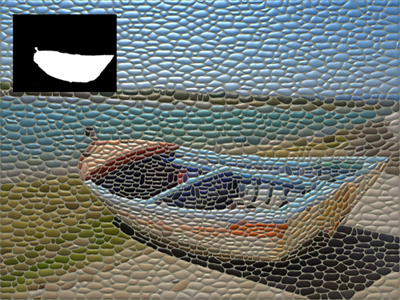}
\end{center}
   \caption{Pebbles under the importance map (inset) are rendered at a higher frequency.}
\label{fig:two_level}
\end{figure}

We can also vary the pebble size through the use of an importance map. The mask in the inset of Figure~\ref{fig:two_level} indicates regions that will be rendered with smaller, more numerous pebbles. This technique is useful for drawing attention to important regions and provides a more detailed representation of the content. 

\subsection{Comparison with related work} \label{compare}

\begin{figure}[b]
\begin{center}
   \includegraphics[width=0.49\linewidth]{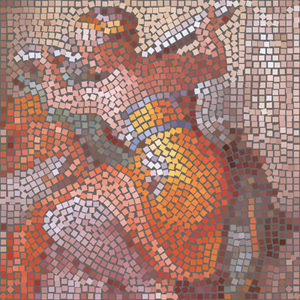}
   \includegraphics[width=0.48\linewidth]{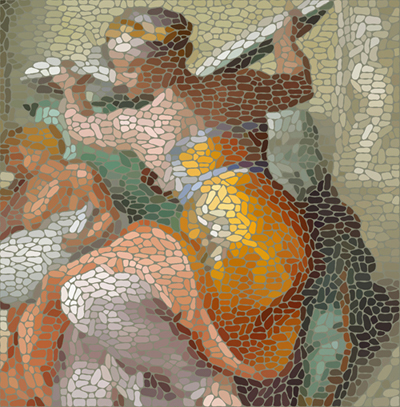}
\end{center}
   \caption{Comparison with Hausner~\cite{Hausner:2001:SDM}. Both results use 2000 tiles. Left: Hausner; right: ours.}
\label{fig:compare_hausner}
\end{figure}

\begin{figure}[b]
\begin{center}
  \includegraphics[width=0.205\linewidth]{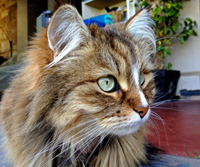}
   \includegraphics[width=0.23\linewidth]{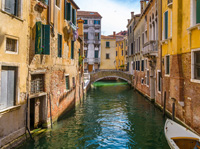}
   \includegraphics[width=0.26\linewidth]{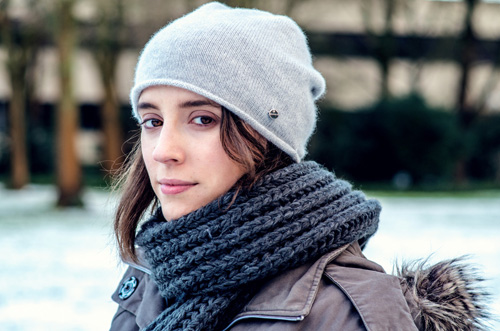}
   \includegraphics[width=0.255\linewidth]{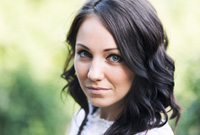} \\[0.05cm]
   \includegraphics[width=0.24\linewidth]{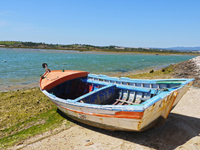}
   \includegraphics[width=0.27\linewidth]{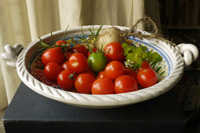}
   \includegraphics[width=0.26\linewidth]{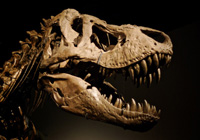}
   \includegraphics[width=0.18\linewidth]{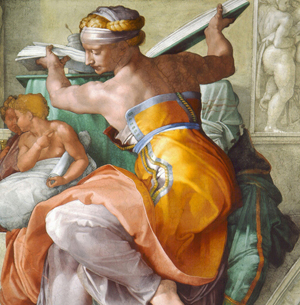}
\end{center}
   \caption{Input images used in Figures~\ref{fig:result}, \ref{fig:size}, \ref{fig:two_level}, and \ref{fig:compare_hausner}.}
\label{fig:input}
\end{figure}

Figure~\ref{fig:compare_hausner} shows a comparison between our method and Hausner's~\cite{Hausner:2001:SDM} using 2000 pebbles. Here, we turn off the lighting effects and make the comparison based on tile shape alone. Note that the color shift between the two examples is due to using different source photographs of the painting. By using heterogeneous shapes, image content can be more accurately portrayed compared to what is possible with an equal number of 2D homogeneous primitives. In our result, the pebble shapes cleanly outline the contours of the figure and its drapery. Where smaller pebbles are needed to fill an image region, our method is not restricted to a uniform pebble size. Both these properties stem from our use of SLIC as the initial segmentation method. Of course, both our method and Hausner's are able to use smaller primitives in regions specified by users.

\begin{figure}
\begin{center}
   \includegraphics[width=1\linewidth]{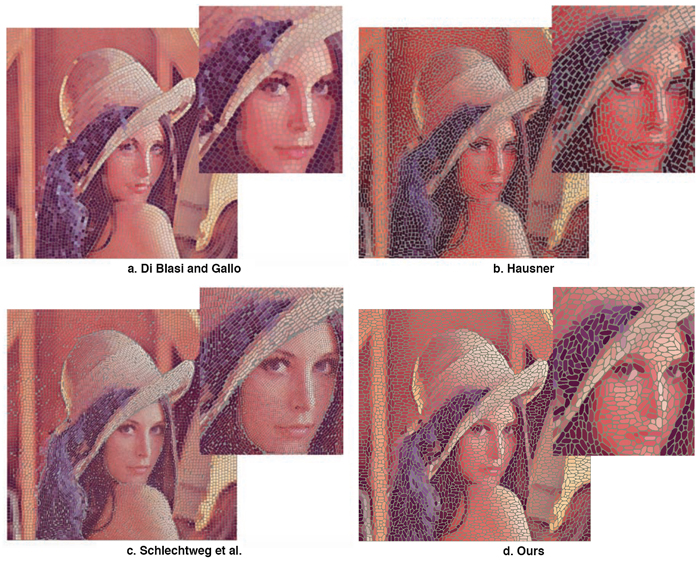}
\end{center}
   \caption{Comparison with previous tile mosaic algorithms.}
\label{fig:comparison}
\end{figure}

Similarly, we compare our method with three previous tile mosaic algorithms on a common image in Figure~\ref{fig:comparison}. Our result is on the bottom right using 3000 pebbles. On the top left Di Blasi and Gallo~\cite{DiBlasi:2005:AM} obtain clean lines and uniform spacing by cutting tiles that overlap perceptual guidelines and neighbouring tiles. The edges in our rendering are obtained through SLIC which adhere well to step edges but fail when perceptual boundaries are not matched with a strong color discontinuity. An example can be seen in the thin strand of feathers above the brim of Lena's hat where pebbles are not constrained to this narrow region. This is a case where perceptual edge detection would be benefit our segmentation. Schlechtweg et al.'s~\cite{Schlechtweg:2005} RenderBots show fine detail by using 9000 primitives but the placement is uneven and the rendering took one hour to complete.

\begin{figure}
\begin{center}
   \includegraphics[width=0.54\linewidth]{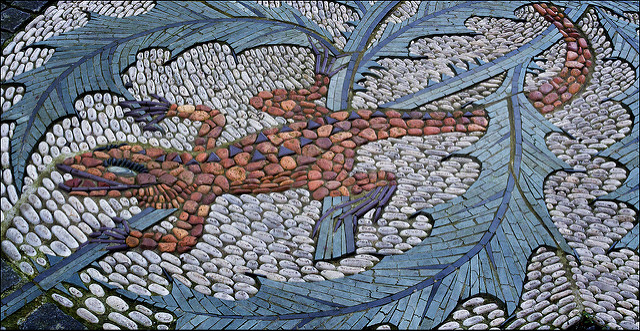}
   \includegraphics[width=0.41\linewidth]{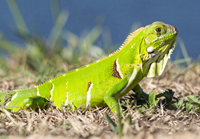} \\[0.05cm]
   \includegraphics[width=0.97\linewidth]{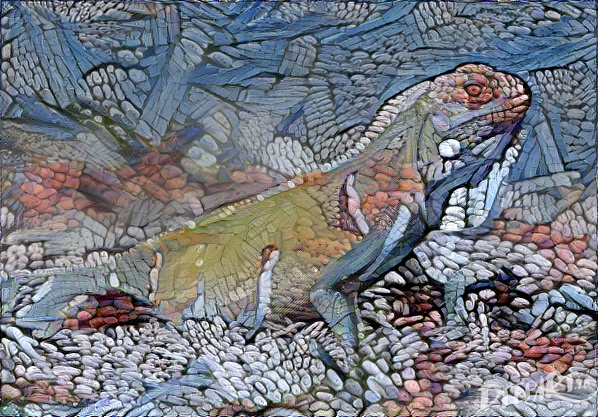} \\[0.05cm]
   \includegraphics[width=0.48\linewidth]{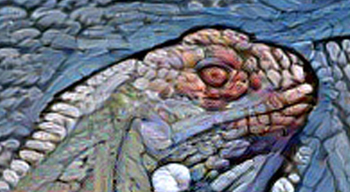}
   \includegraphics[width=0.48\linewidth]{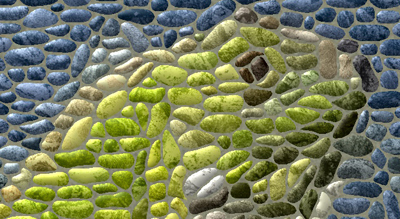}
\end{center}
   \caption{Comparison with neural style. Top left: style example; top right: input image; center: pebble mosaic rendered with neural style~\cite{Gatys:2016} as implemented at {\em deepart.io}; bottom left: detail; bottom right: detail of Figure~\ref{fig:result} (bottom).}
\label{fig:neural_style}
\end{figure}

Recently, there has been a lot of attention to using convolutional neural networks for image stylization~\cite{Gatys:2016,Johnson:2016}. In the top left of Figure~\ref{fig:neural_style} we show a result obtained from {\em deepart.io}, a popular online implementation of Gatys et al.'s method~\cite{Gatys:2016}. The high-level semantic features used in neural style transfer preserve image features better than the low-level color features that we use; compare the detail images on the bottom left to the bottom right of Figure~\ref{fig:neural_style}. The iguana's eye clearly highlights the advantage of using semantic features: style transfer reproduced the eye using a single pebble, improving recognizability. Our method, in contrast, uses a number of pebbles that is dependent on the SLIC super-pixel size; it artificially breaks the eye into three pebbles. The advantage of our method lies in explicitly modeling pebble shapes. The texture that is produced by neural style transfer (center) only roughly approximates what is found in the style example shown at the top left. For example, the definition of individual pebbles is completely lost in parts of the background and the side of the iguana's head. In contrast, our method explicitly models individual pebble geometry and can output well-defined shapes at any resolution.

\begin{figure}
\begin{center}
   \includegraphics[width=0.48\linewidth]{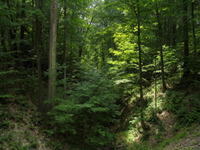}
   \includegraphics[width=0.48\linewidth]{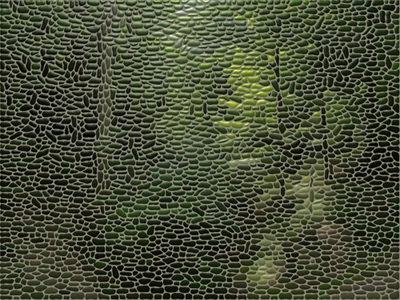} \\[0.05cm]
   \includegraphics[width=0.48\linewidth]{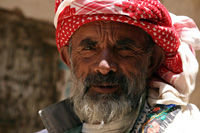}
   \includegraphics[width=0.48\linewidth]{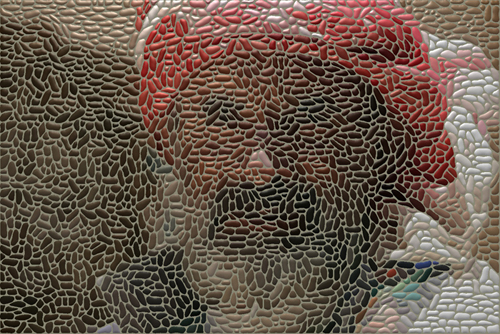}
\end{center}
   \caption{Limitations of our method. Top: high-frequency features; bottom: semantic content.}
\label{fig:limitations}
\end{figure}

\subsection{Limitations}

Our method performs best on images with high contrast and clear distinctions between regions of differing semantic content. Due to the relatively large scale of the pebbles, some subtle image features or tiny details can be lost. Figure~\ref{fig:limitations} (top) shows an image dominated by high-frequency content. In our rendering on the right, only a large-scale impression of the scene is captured. Reducing pebble size is only a limited option since, past a certain scale, the cement between the pebbles will feature as prominently as the pebbles themselves. On the bottom of Figure~\ref{fig:limitations} the facial features are poorly represented. SLIC does not effectively cope with the lighting changes in the area of the man's noise, for example. Either more sophisticated low-level processing or learning-based semantic segmentations could improve on our results, and both are promising directions for future work.

\begin{figure}
\begin{center}
   \includegraphics[width=0.48\linewidth]{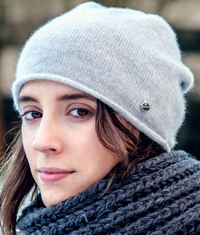}
   \includegraphics[width=0.48\linewidth]{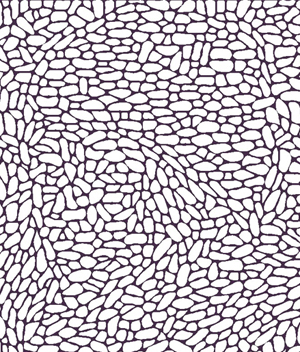}
\end{center}
   \caption{Pebble layout without colorization.}
\label{fig:no_color}
\end{figure}

Continuing our discussion on color, we also note that our resulting images would be difficult to recognize based on pebble layout only. See Figure~\ref{fig:no_color} for an example of a black and white pebble layout. Without colorization, the orientation and pebble boundaries can only hint at the underlying image. More work could be done to emphasize the structural content of the image by varying pebble shape and size, linking size and shape variation to image content instead of varying pebble dimensions with random factors. At the same time, it might be possible to improve our pebble colors. Because we add lighting effects to a base color derived from the image, the final pebble color distribution is not necessarily very close to the desired color. We would be able to improve the mosaic with better integration of the lighting process and the selection of base color. 

Processing time is also an issue. Our main bottleneck is solving the numerous matrices that construct the heightfield. Taking advantage of parallelization would help. Also, solving at a lower resolution and smoothing the results could improve timing. 

Although we think that smooth river-worn pebbles are the most common type for pebble mosaics, more varied rock types in principle could be used, and this paper did not attempt to treat these.

\section{Conclusion} \label{con}

In this paper we present a method to render 3D pebble mosaics. Digital mosaics have been presented in the NPR literature previously, but only in the context of tiling a 2D surface; here, we not only create a tiling representing pebbles, but also generate a heightfield for the pebbles so that they can be rendered.

 Our method starts by segmenting the image plane with SLIC, equipped with a modified distance metric. The resulting super-pixels adhere to image boundaries and hence no further edge detection is required. By varying the size, orientation, and aspect ratio of the super-pixels, we obtain pebble shapes that are highly expressive in their depiction of image content. 

We construct the geometry of each pebble by solving a Laplace equation on the domain between two contours. The resulting heightfield can then be rendered using a variety of lighting techniques beyond the simple Phong shading model we use in this paper. In addition, since we have synthesized 3D geometry, our pebble mosaics can be used in novel applications, from 3D virtual environments to physical 3D printed objects.

In the future we would like to use semantic segmentation to improve the initial super-pixel clustering. Important image regions, especially on the human face, could benefit by constraining clustering to regions of similar content. Better use of low-level image features could improve on the SLIC segmentation. Pebble texture could also be customized to suggest image details at a scale below the size of individual pebbles. This addition would bridge the gap between tile and multi-picture mosaics, as defined by Battiato et al.~\cite{Battiato:2007}, and strengthen the connection between the original image and its mosaic representation.


{\small

}

\end{document}